\documentclass[paper]{sigirforum}

\def\pubissue{Vol. 56 No. 1 June 2022}

\begin{document}
\title{Responsible Artificial Intelligence -- from Principles to Practice}

\authors{
\author[virginia@cs.umu.se]{Virginia Dignum}{Ume{\aa} University}{Sweden}
}

\maketitle 
\begin{abstract}
The impact of Artificial Intelligence does not depend only on fundamental research and technological developments, but for a large part on how these systems are introduced into society and used in everyday situations. AI is changing the way we work, live and solve challenges but concerns about fairness, transparency or privacy are also growing.
Ensuring responsible, ethical AI is more than designing systems whose result can be trusted. It is about the way we design them, why we design them, and who is involved in designing them. In order to develop and use AI responsibly, we need to work towards technical, societal, institutional and legal methods and tools which provide concrete support to AI practitioners, as well as awareness and training to enable participation of all, to ensure the alignment of AI systems with our societies’ principles and values. \\
This paper is a curated version of my keynote at the Web Conference 2022.
\end{abstract}

\section{Introduction}
Ensuring the responsible development and use of AI is becoming a main direction in AI research and practice. Governments, corporations and international organisations alike are coming forward with proposals and declarations of their commitment to an accountable, responsible, transparent approach to AI, where human values and ethical principles are leading. 

Currently, there are over 600 AI-related policy recommendations, guidelines or strategy reports, which have been released by prominent intergovernmental organisations, professional bodies, national-level committees and other public organisations, non-governmental, and private for-profit companies. A recent study of the global landscape of AI ethics guidelines shows that there is a global convergence around five ethical principles: Transparency, Justice and Fairness, Non-Maleficence, Responsibility, and Privacy \citep{jobin2019a}. 
These are much-needed efforts, but still much work is needed to ensure that all AI is developed and used in responsible ways that contribute to trust and well-being. Nevertheless, even though organisations agree on the need to consider ethical, legal and societal principles, how these are interpreted and applied in practice, varies significantly across the different recommendation documents. 

At the same time, the growing hype around `AI’ is blurring its definition and shoving into the same heap concepts and applications of many different sorts. A hard needed first step in the responsible development and use of AI is to ensure a proper AI narrative, one that demystifies its capabilities, minimises both overselling and underselling of AI-driven solutions, and 
that enables wide and inclusive participation in the discussion on the role of AI in society. Understanding the capabilities and addressing the risks of AI, requires that all of us, from developers to policy-makers, from provides to end-users and bystanders, have a clear understanding of what AI is, how it is applied and what are the opportunities and risks involved.

\section{What AI is not: data, algorithms, magic}
Without a proper understanding of what AI is and what it can, and cannot, do, all the efforts towards governance, regulation and the responsible development and use of AI have the risk to become void. Current AI narratives bring forward benefits and risks and describe AI in many different ways, from the obvious next step in digitisation to some kind of magic. If the `business as usual' narrative is detrimental of innovation and contributes to the maintenance current power structures, the `magic' narrative, well fed by science fiction and the popular press, often supports a feeling that nothing can be done against such an all-knowing entity that rules over us in possibly unexpected ways, either solving all our problems, or destroying the world in the process. In both cases, the danger is that the message is that little can be done against the risks and challenges of AI.



Currently, AI 
is mostly associated with 
data-driven techniques that use statistical methods to enable computers to perceive some characteristics of their environment. Such techniques are particularly efficient in perceiving images, written or spoken text, as well as the many applications of structured data. These techniques are extremely successful at pattern matching: By analysing many thousands of examples (typically a few million), the system is able to identify commonalities in these examples, which then enable it to interpret data that it has never seen before, which is often referred to as prediction. These results, however impressive and useful, are still far from any thing that we would consider as `intelligent'. Moreover, data-driven approaches to AI have been proven to be problematic in terms of bias, explanation, inclusion and transparency. Algorithms are too complex for human inspection, and a over-reliance on data, condemns the future to repeat the past. Indeed, data is always about the past, and decisions on which, how, when, why collect and maintain data fundamentally influence the availability and quality of data. Those that have the power to decide on data, have the power to determine how AI system will be design, deployed and used.\\

AI is based on algorithms, but then so is any computer program and most of the technologies around us. Nevertheless, the concept of `algorithm’ is achieving magical proportions, used right and left to signify many things, \textit{de facto} seen as a synonym to AI. 
The easiest way to understand an algorithm is as a recipe, a set of precise rules to achieve a certain result. Every time you multiply two numbers, you are using an algorithm, as well as you are when you are baking an apple pie. However, by itself, the recipe has never turned into an apple pie; and, the end result of your pie has as much to do with your baking skills and your choice of ingredients, as with the choice for a specific recipe. The same applies to AI algorithms: for a large part the behaviour and results of the system depends on its input data, and on the choices made by those that developed, trained and selected the algorithm. In the same way as we have the choice to use organic apples to make our pie, in AI we also must consider our choices on which models, data to use, who to include in the design and considerations about impact, and how these choices respect and ensure fairness, privacy, transparency and all other values we hold dear. 

\subsection{AI is a socio-technical ecosystem: datification, power and costs}
If AI is not intelligent, nor magic, nor business as usual, nor an algorithm, how best can we describe AI in order to take into account not only its capabilities but also its societal implications?
AI is first and foremost technology that can automatise (simple, lesser) tasks and decision making processes. At the present, AI systems are largely incapable of understanding meaning and the context of their operation and results. 
At the same time, considering its societal impact and need for human contribution, AI is much more than an automation technique. When considering effects and the governance thereof, the technology, or the artefact that embeds that technology, cannot be separated from the ecosystem of which it is a component. 
In this sense, AI can best be understood as a socio-technical ecosystem, recognising the interaction between people and technology, and how complex infrastructures affect and are affected by society and by human behaviour \citep{raibook2019}. 
AI is not just about the automation of decisions and actions, the adaptability to learn from the changes affected in the environment, and the interactivity required to be sensitive to the actions and aims of other agents in that environment, and decide when to cooperate or to compete. It is mostly about the structures of power, participation and access to technology that determine who can influence which decisions or actions are being automated, which data, knowledge and resources are used to learn from, and how interactions between those that decide and those that are impacted are defined and maintained. 

Much has been said about the dangers of biased data, and discriminating applications. Minimising or eliminating discriminatory bias or unfair outcomes is more than excluding the use of low-quality data. The design of any artefact, such as an AI system, is in itself an accumulation of choices and choices are biased by nature as they involve selecting an option over another. Most importantly, it starts with the current reliance on data as a measure of what can be done. Increasingly, we are seeing that the availability of that (or the possibility to access data) is taken as a guiding criteria to solving societal issues. If there is data, it is a problem we can address, but if there is no data, there is no problem. This is intrinsically related to power and to power structures. Those that can decide on which problems are worth address, are shaping not only how AI is being developed and used, which technologies to use and what values to prioritise. Those in power are shaping the way we live with AI and how our future societies will look like. 

Nevertheless, attention for the societal, environmental and climate costs of AI systems is increasing. All these must be included in any effort to ensure the responsible development and use of AI. 
A responsible, ethical, approach to AI will ensure transparency about how adaptation is done, responsibility for the level of automation on which the system is able to reason, and accountability for the results and the principles that guide its interactions with others, most importantly with people. In addition, and above all, a responsible approach to AI makes clear that AI systems are artefacts manufactured by people for some purpose, and that those which make these have the power to decide on the use of AI. It is time to discuss how power structures determine AI and how AI establishes and maintains power structures, and on the balance between, those who benefit from, and those who are harmed by the use of AI \citep{crawford2021a}. 

\section{Responsible AI - The question zero}
Responsible AI (or Ethical AI, or Trustworthy AI) is not, as some may claim, a way to give machines some kind of `responsibility’ for their actions and decisions, and in the process discharge people and organisations of their responsibility. On the contrary, responsible development and use of AI requires more responsibility and more accountability from the people and organisations involved: for the decisions and actions of the AI applications, and for their own decision of using AI in a given application context \citep{dignum22}. When considering effects and the governance thereof, the technology, or the artefact that embeds that technology, cannot be separated from the socio-technical ecosystem of which it is a component. Guidelines, principles and strategies to ensure trust and responsibility in AI, must be directed towards the socio-technical ecosystem in which AI is developed and used. It is not the AI artefact or application that needs to be ethical, trustworthy, or responsible. Rather, it is the social component of this ecosystem that can and should take responsibility and act in consideration of an ethical framework such that the overall system can be trusted by the society. Having said this, governance can be achieved by several means, softer or harder. Currently several directions are being explored, the main ones are highlighted in the remainder of this section. Future research and experience will identify which approaches are the most suitable, but given the complexity of the problem, it is very likely that a combination of approaches will be needed. 

Responsible AI is more than the ticking of some ethical `boxes’ or the development of some add-on features in AI systems. Nevertheless, developers and users can benefit from support and concrete steps to understand the relevant legal and ethical standards and considerations when making decisions on the use of AI applications. Impact assessment tools provide a step-by-step evaluation of the impact of systems, methods or tools on aspects such as privacy, transparency, explanation, bias, or liability \citep{taddeo2018a}. 

Inclusion and diversity are a broader societal challenge central to AI development. It is therefore important that as broad a group of people as possible have a basic knowledge of AI, what can (and can’t) be done with AI, and how AI impacts individual decisions and shapes society. 
In parallel, research and development of AI systems must be informed by diversity, in all the meanings of diversity, and obviously including gender, cultural background, and ethnicity. Moreover, AI is not any longer an engineering discipline and at the same time there is growing evidence that cognitive diversity contributes to better decision making. Therefore, it is important to diversify the discipline background and expertise of those working on AI to include AI professionals with knowledge of, amongst others, philosophy, social science, law and economy. 

\section{Design for Responsibility}
A multidisciplinary stance supporting understanding and critiquing the intended and unforeseen, positive and negative, and the socio-political consequences of AI for society, is core to the responsible design of AI systems. This multidisciplinary approach is fundamental to understand governance, not only in terms of competences and responsibilities, but also in terms of power, trust and accountability; to analyse the societal, legal and economic functioning of socio-technical systems, providing value-based design approaches and ethical frameworks for inclusion and diversity in design, and how such strategies may inform processes and results.

Achieving trustworthy AI systems is a multifaceted complex process, which requires both technical and socio-legal initiatives and solutions to ensure that we always align an intelligent system’s goals with human values. Core values, as well as the processes used for value elicitation, must be made explicit and that all stakeholders are involved in this process. Furthermore, the methods used for the elicitation processes and the decisions of who is involved in the value identification process must be clearly identified and documented.

Where it concerns the design process itself, responsibility includes the need to elicit and represent stakeholders, their values and expectations, as well as ensuring transparency about how such values are interpreted and prioritised in the concrete functionalities of the AI system. Design for Values methodologies \citep{Hoven05,friedman2006} are often used for this end, providing a structured way for translation from abstract values into concrete norms comprehensive enough so that fulfilling the norm will be considered as adhering to the value. Following a Design for Values approach, the shift from abstract to concrete necessarily involves careful consideration of the context.
Design for Values approach means that the process needs to include activities for (i) the identification of societal values, (ii) deciding on a moral deliberation approach (e.g. through algorithms, user control or regulation), and
 (3) linking values to formal system requirements  and concrete functionalities.

My research group is developing the Glass Box framework \citep{aler2019glass} that is both an approach to software development, a verification method and a source of high-level transparency for intelligent systems. It provides a modular approach integrating verification with value-based design.

\section{Conclusions}
Increasingly, AI systems will be taking decisions that affect our lives, in smaller or larger ways. In all areas of application, AI must be able to take into account societal values, moral and ethical considerations, weigh the respective priorities of values held by different stakeholders and in multicultural contexts, explain its reasoning, and guarantee transparency. As the capabilities for autonomous decision-making grow, perhaps the most important issue to consider is the need to rethink responsibility. Being fundamentally tools, AI systems are fully under the control and responsibility of their owners or users. However, their potential autonomy and capability to learn, require that design considers accountability, responsibility and transparency principles in an explicit and systematic manner. The development of AI algorithms has so far been led by the goal of improving performance, leading to opaque black boxes. Putting human values at the core of AI systems calls for a mind-shift of researchers and developers towards the goal of improving transparency rather than performance, which will lead to novel and exciting techniques and applications. 

Finally, it is crucial to understand responsibility, regulation and ethics as stepping-stone for innovation, rather than the often referred hinder to innovation. True innovation is moving technology forward, not use existing technology ‘as is’. Taking responsibility and regulation as beacons pointing the direction to move, will not only lead to better technology but also ensure trust and public acceptance, serve as a drive for transformation and for business differentiation. Efforts in fundamental research are part of this. Currently, much AI `innovation' is based on brute force: when the main data analytics paradigm is correlation, better accuracy is achieved by increased amount of data and computational power. The effort/accuracy ratio is huge. However, human intelligence is not based on correlation, and includes causality and abstraction. 
Responsibility in AI is not just about ethics, bias, and trolley problems. It is also about responsible innovation: ensuring the best tools for the job, minimize side effects. Responsible innovation in AI requires ``a shift from a perspective in which learning is more or less the only first-class citizen to one in which learning is a central member of a broader coalition that is more welcoming to variables, prior knowledge, reasoning, and rich cognitive models."\citep{marcus20}.

\subsubsection*{Acknowledgements}
This work was partially supported by the Wallenberg AI, Autonomous Systems and Software Program (WASP), funded by the Knut and Alice Wallenberg Foundation and by the European Commission’s Horizon2020 project HumaneAI-Net (grant 952026).

\bibliography{refs}

\end{document}